\RequirePackage[l2tabu, orthodox]{nag}
\RequirePackage{ifpdf}
\RequirePackage[usenames,dvipsnames]{xcolor}
\RequirePackage{tikz}

\documentclass[prl,twocolumn,preprintnumbers,footinbib]{revtex4-1}%

\usepackage{hyperref}
\usepackage[utf8]{inputenc}%
\usepackage{mathtools, amscd, dsfont, amsthm, amssymb,array}
\usepackage{graphicx}
\usepackage[toc,page]{appendix}
\usepackage{etoolbox}
\mathtoolsset{showonlyrefs=false}
\usetikzlibrary{arrows,calc,shapes.misc,decorations.markings,snakes}
\tikzset{cross/.style={cross out, draw=black, minimum size=2*(#1-\pgflinewidth), inner sep=0pt,
        outer sep=0pt},cross/.default={3pt},
    gluon/.style={decorate, decoration={coil,aspect=0.9,segment length=5pt, amplitude=3pt}}}

\usepackage[english]{babel}
\usepackage{epsf,epsfig}
\usepackage{amssymb,amsmath,amsfonts}
\usepackage{mathrsfs}
\usepackage{color}
\usepackage{enumitem}

\usepackage{tikz}

\usepackage[mathscr]{eucal}

\def\cO {\mathcal{O}}

\def\cK{\mathcal{K}}

\def\cN{\mathcal{N}}
\def\cF{\mathcal{F}}

\def\cR{\mathcal{R}}

\def\zb{\bar{z}}

\newcommand{\be}{\begin{equation}}
\newcommand{\ee}{\end{equation}}
\newcommand{\bea}{\begin{eqnarray}}
\newcommand{\eea}{\end{eqnarray}}

\begin{document}
\begin{flushright}
\preprint{DESY 19-180, CERN-TH-2019-204, IPPP/19/86, UUITP-48/19, SLAC-PUB-17491}
\end{flushright}

\title{All-order amplitudes at any multiplicity in the multi-Regge limit}

\author{V. Del Duca${}^{1,2}$, S. Druc${}^3$, J. M. Drummond${}^3$, C. Duhr${}^{4}$, F. Dulat${}^5$, R. Marzucca${}^6$, G. Papathanasiou${}^7$, B. Verbeek${}^8$}
\affiliation{
$\rule{0pt}{.01cm}$
\makebox[\textwidth][c]{${}^{1}$Institute for Theoretical Physics, ETH Z\"urich, 8093 Z\"urich, Switzerland}\\
\makebox[\textwidth][c]{${}^{2}$INFN, Laboratori Nazionali di Frascati, 00044 Frascati (RM), Italy}\\
\makebox[\textwidth][c]{${}^{3}$School of Physics and Astronomy, University of Southampton, Highfield, SO17 1BJ, United Kingdom}\\
\makebox[\textwidth][c]{${}^{4}${}Theoretical Physics Department, CERN, CH-1211 Geneva 23, Switzerland}\\
\makebox[\textwidth][c]{${}^5$SLAC National Accelerator Laboratory, Stanford University,  Stanford, CA 94309, USA}\\
\makebox[\textwidth][c]{${}^6$IPPP, Department of Physics, Durham University, Durham DH1 3LE, United Kingdom}\\
\makebox[\textwidth][c]{${}^7$DESY Theory Group, DESY Hamburg, Notkestra{\ss}e 85, D-22607 Hamburg, Germany}\\
\makebox[\textwidth][c]{${}^8$Department of Physics and Astronomy, Uppsala University, 75108 Uppsala, Sweden}
}

\begin{abstract}
\noindent
We propose an all-loop expression for scattering amplitudes in planar $\mathcal{N}=4$ super Yang-Mills theory in multi-Regge kinematics valid for all multiplicities, all helicity configurations and arbitrary logarithmic accuracy. Our expression is arrived at from comparing explicit perturbative results with general expectations from the integrable structure of a closely related collinear limit.
A crucial ingredient of the analysis is an all-order extension for the central emission vertex that we recently computed at next-to-leading logarithmic accuracy. As an application, we use our all-order formula to prove that all amplitudes in this theory in multi-Regge kinematics are single-valued multiple polylogarithms of uniform transcendental weight.
\end{abstract}
\maketitle



Recent years have seen tremendous progress in our understanding of multi-loop multi-leg scattering amplitudes in planar $\cN=4$ super Yang-Mills (SYM) theory. Its $S$-matrix exhibits a hidden dual conformal (DC) symmetry~\cite{Drummond:2008vq}, which closes with the ordinary conformal symmetry into a Yangian algebra \cite{Drummond:2009fd}.

The DC symmetry is broken by infrared (IR) divergences. Such divergences are universal and independent of the hard scattering process and it is possible to construct DC-invariant functions by considering ratios where all IR-divergences cancel. We denote by $\cR_N$ the IR-finite ratio of the $N$-point color-ordered amplitude and the Bern-Dixon-Smirnov (BDS) ansatz~\cite{Bern:2005iz}, defined (loosely) as the exponential of the one-loop amplitude multiplied by the cusp anomalous dimension $\Gamma_{\rm cusp}$ \cite{Beisert:2006ez}. DC-invariance dictates that $\cR_N$ only depends on $3N-15$ independent cross-ratios. In particular, $\cR_N$ is trivial for $N\le5$~\cite{Drummond:2007au}, and is known analytically in general kinematics for
$N=6$ through seven loops~\cite{DelDuca:2009au,DelDuca:2010zg,Goncharov:2010jf,Dixon:2011pw,Dixon:2011nj,Dixon:2013eka,Dixon:2014voa,Dixon:2014xca,Dixon:2014iba,Dixon:2015iva,Caron-Huot:2016owq,Caron-Huot:2019vjl},
and for $N=7$ through four loops~\cite{Golden:2014xqf,Golden:2014xqa,Drummond:2014ffa,Dixon:2016nkn,Drummond:2018caf}, at the level of the symbol \cite{Goncharov:2010jf}.

Explicit data for small $N$ reveals that the perturbative expansion of $\cR_N$ can often be expressed in terms of a class of iterated integrals known as \emph{multiple polylogarithms} (MPLs)~\cite{Goncharov:2001}. Moreover only MPLs of (transcendental) weight $2L$ contribute to an $L$-loop amplitude, where weight is the number of iterated integrations.

\begin{figure}[h]
        \center
        \epsfig{file=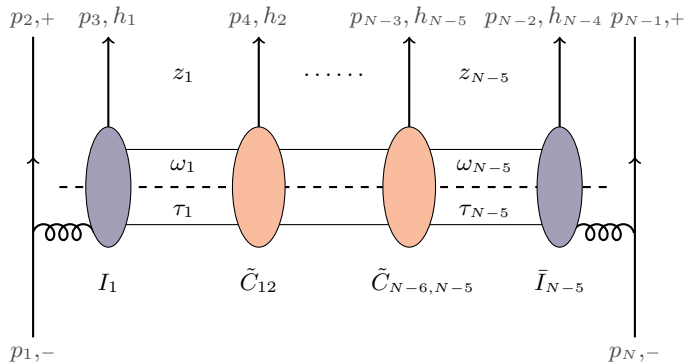} 
        \caption{Fourier-Mellin factorisation of $2\to N-2$ gluon amplitude in multi-Regge kinematics.}
        \label{fig:bfklViz}
    \end{figure}

The mathematical beauty and simplicity of the available perturbative results hint at some deeper structure governing amplitudes in planar $\cN=4$ SYM theory. This is corroborated by the fact that infinite-dimensional symmetries, like the Yangian symmetry of $\cN=4$ SYM, are a hallmark of integrability. One should then be able to compute $\cR_N$ at any value of the coupling. A major step in this direction was taken in~\cite{Basso:2013vsa,Basso:2013aha,Basso:2014koa,Basso:2014nra,Basso:2014pla}, where it was argued that amplitudes (or their dual Wilson loops~\cite{Alday:2007hr,Drummond:2007aua,Brandhuber:2007yx,Bern:2008ap,Drummond:2008aq}) can be computed through an integrable flux-tube picture. The dream of computing amplitudes analytically at any value of the coupling constant $g^2$, or at least at any order in perturbation theory, has not yet been achieved.

Here we present for the first time a way to compute scattering amplitudes in planar $\cN=4$ SYM to any order in the coupling, for any helicity configuration and any number of external legs, albeit in the simplified kinematic setup of multi-Regge kinematics (MRK) where the produced particles are strongly ordered in rapidity and have comparable transverse momenta. While in Euclidean kinematics the ratios $\cR_N$ become trivial in the
limit~\cite{Brower:2008nm,Bartels:2008ce,DelDuca:2008pj,Bartels:2008sc,Brower:2008ia,DelDuca:2008jg}, they develop a non-trivial kinematic dependence when some of the energies of the produced gluons are analytically continued to negative values~\cite{Bartels:2008ce,Bartels:2008sc}. Here we focus on the situation where all the centrally-produced gluons have a negative energy, and we propose a formula for any amplitude in MRK in this theory.

\section{The $N$-particle dispersion integral}

In MRK a subset of $N-5$ cross-ratios, denoted by $\tau_i$, approach zero. $\cR_N$ can then be expressed at each order as a polynomial in large logarithms $\log \tau_i$, multiplied by functions of the $2N-10$ remaining real degrees of freedom. The latter are conveniently described by $N-5$ complex variables $z_i$, see \cite{DelDuca:2016lad} and references therein for these standard conventions.
We conjecture that, to all orders, $\cR_N$ can be written as a Fourier-Mellin (FM) integral with a factorised form, as also depicted in fig.~\ref{fig:bfklViz},
\begin{align}
\label{conj}
\frac{\mathcal{R}_N e^{i \Gamma \delta}}{2\pi i}\!=\! &\prod_{r=1}^{N-5} \biggl[\sum_{n_r } \Bigl(\frac{z_r}{\bar{z}_r}\Bigr)^{\frac{n_r}{2}}\!  \!\!\! \int_{\mathcal{C}} \frac{d \nu_r}{2\pi} \frac{|z_r|^{2i \nu_r} \tilde{\Phi}_r}{(-\tau_r+i0)^{\omega_r}} \biggr] \notag \\
& \,\, \times I_1^{h_1} \tilde{C}_{12}^{h_2} \ldots \tilde{C}^{h_{N-5}}_{N-6,N-5} \bar{I}^{h_{N-4}}_{N-5} \,.
\end{align}
Equation~\eqref{conj} extends similar formulas in the literature for restricted subsets of amplitudes at leading logarithmic accuracy (LLA) and beyond~\cite{Bartels:2008sc, Lipatov:2010qf,Fadin:2011we,Bartels:2011ge,Lipatov:2012gk,Bartels:2013jna,DelDuca:2016lad,DelDuca:2018hrv}, see also \cite{Marzucca:2018ydt} for an application.
The ratio $\cR_N$  depends on the helicities $h_r$ of all centrally-produced particles.
The building blocks of the integrand $\omega_r$, $\tilde{\Phi}_r$, $I_r$ and  $\tilde{C}^{h_{r+1}}_{r,r+1}$ are known as the Balitsky-Fadin-Kuraev-Lipatov (BFKL) eigenvalue, impact factor product, helicity flip kernel and (rescaled) central emission block (see aforementioned references, and references therein). They are functions of the FM variables $(\nu_r,n_r)$, whose precise form will be presented below, and we use a shorthand notation $\omega_r = \omega(\nu_r,n_r)$ and $\tilde{C}^{h_{r+1}}_{r,r+1} = \tilde{C}^{h_{r+1}}(\nu_r,n_r,\nu_{r+1},n_{r+1})$ etc.
The phase $e^{i  \Gamma \delta}$, where $\Gamma \equiv \Gamma_{\rm cusp}/4$, captures terms in the BDS ansatz that do not vanish after analytic continuation in MRK \footnote{Explicitly $\delta(z) =
\pi \log\frac{|\rho_1|^2}{|1-\rho_1|^2|1-\rho_{N-5}|^2}$ where the $\rho_i$ are defined in terms of the $z_i$ via $z_i=\frac{(1-\rho_{i+1})(\rho_i-\rho_{i-1})}{(1-\rho_{i-1})(\rho_i-\rho_{i+1})}$ together with $\rho_0=0$ and $\rho_{N-4}=\infty$, see e.g. \cite{Lipatov:2010qf,Bartels:2011ge,Bartels:2013jna,DelDuca:2018raq}}.

\begin{figure}[!t]
        \center
        \epsfig{file=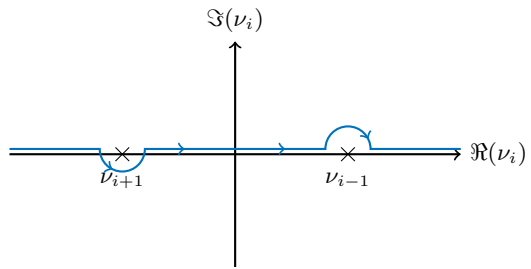} 
        \caption{Contour of integration $\mathcal{C}$ for the integral \eqref{conj}, with $\nu_{N-4}=-\pi \Gamma$, $\nu_{0}=\pi \Gamma$ corresponding to the boundary cases.
        \label{fig:contourOne}}
    \end{figure}

In the limit where one of the centrally-produced gluons becomes soft, $\cR_N$ should reduce to $\cR_{N-1}$. Provided that the building blocks have at most simple poles on the integration axis, this then dictates that the contour $\mathcal{C}$ must take the form shown in fig.~\ref{fig:contourOne}, and implies the following \emph{exact bootstrap conditions}~\cite{Caron-Huot:2013fea,DelDuca:2018hrv},
\begin{align}
\omega(\pm \pi \Gamma,0) = 0, \quad  \mathrm{Res}&_{\nu=\pm\pi \Gamma}\left(\tilde{\Phi}(\nu,0)\right)=\pm\frac{1}{2\pi }\,,\\
\tilde C^h(\pi \Gamma,0,\nu_2,n_2) &=2\pi i  \,I^h(\nu_2,n_2)\,,
\\
\tilde C^h(\nu_1,n_1,-\pi \Gamma,0)&=-2\pi i \,\bar I^h(\nu_1,n_1)\,,\label{eq:R7hhh_bootstrapcond2}\\
\underset{{\nu_1=\nu_2}}{\mathrm{Res}}\tilde C^h(\nu_1,n_2,\nu_2,n_2)&=\frac{-i (-1)^{n_2}e^{i\pi\omega(\nu_2,n_2)}}{\tilde \Phi(\nu_2,n_2)}\,,
\label{eq:R7hhh_bootstrapcond3}
\\
\tilde C^h(-\pi \Gamma,0,\nu_2,n_2)&=\tilde C^h(\nu_1,n_1,\pi \Gamma,0)=0\,.
\end{align}

Let us now proceed to fully specify the integral (\ref{conj}), by providing explicit expressions for its building blocks. The BFKL eigenvalue $\omega_r$, impact factor product $\tilde{\Phi}_r$ and helicity flip kernel $I_r$ have already been determined to all loops~\cite{Basso:2014pla}, by means of an analytic continuation from the collinear limit. The latter limit is also described by a dispersion integral very similar to (\ref{conj}), whose building blocks are governed by an integrable flux tube, and may thus be computed at finite coupling within the Pentagon operator product expansion (OPE) \cite{Basso:2013vsa,Basso:2013aha,Basso:2014koa,Basso:2014nra} approach. Then, the authors of \cite{Basso:2014pla} were able to connect the multi-Regge and collinear integrands by analytically continuing in the integration variable, and in particular obtain $\omega_r,\tilde{\Phi}_r$ and $I_r$ from their OPE counterparts, the gluonic excitation energy, measure and next-to-maximally helicity violating (NMHV) impact factor respectively. A feature of this analysis is that at finite coupling it is more natural to use rapidities $u_r$ rather than $\nu_r$ as integration variables, giving rise to the following implicit all-loop dispersion relation,
\be\label{eq:nutou_weak}
\nu_r = u_r - 2 g (\mathbb{Q} \cdot \mathbb{M} \cdot \tilde{\kappa})_1\,,\quad  \omega_r= -4g(\mathbb{Q} \cdot \mathbb{M} \cdot \kappa)_1\,.
\ee
The sources $\kappa$ and $\tilde{\kappa}$ are infinite-dimensional vectors and are described explicitly in the Appendix along with the matrices $\mathbb{Q}$ and $\mathbb{M}$ which essentially encode the Beisert-Eden-Staudacher kernel \cite{Beisert:2006ez,Benna:2006nd}. The subscript 1 in (\ref{eq:nutou_weak}) means the first component of the vector.

\section{Central emission vertex}
The only quantity in~\eqref{conj} only known at leading order (LO)~\cite{Bartels:2011ge} and next-to-LO~\cite{DelDuca:2018hrv} is the central emission vertex $C^{\pm}_{r,r+1}$.
A main result of this paper is a conjecture for $C^{\pm}_{r,r+1}$ to all orders in the coupling, as we now move on to describe. We focus on the vertex for the emission of a positive helicity gluon. The case of negative helicity is then recovered from the helicity flip kernel \footnote{Explcitly the helicity-flip kernels are $I^+_r=1$ and $I^{-}_r = x^-_r/x^+_r$, where the Zhukowski variables $x^\pm_r$ are given in (\ref{Zhukowski})},
\be
\tilde C_{r,r+1}^{-} = \tilde C_{r,r+1}^{+} \bar{I}^{-}_r I^{-}_{r+1}\,.
\ee

Our analysis parallels that of~\cite{Basso:2014pla} for $N=6$. We assume that also for $N=7$, the dispersion integral~\eqref{conj} can be obtained by analytically continuing the contribution of gluon excitations to the pentagon OPE through the branch cut at $u_r=-i n_r/2\pm 2g$ in the rapidity plane. It follows that the central emission vertex is the analytic continuation of the new OPE building block appearing at this multiplicity, known as the \emph{gluon pentagon transition}~\cite{Basso:2014nra}. Performing the analytic continuation in full generality is quite complicated, but we are able to present a conjectural all-orders form for the central emission vertex by continuing certain factors of the pentagon transition, and fixing the remaining proportionality coefficient by consistency with known perturbative data in MRK.
More precisely, our conjecture reads
\be\label{eq:C_all_loop}
 \tilde C^+_{12}=\frac{\tilde C^{(0)}_{12}}{g^2} k_{12} Z_{12} \exp(f_{ 1  2}-f_{ \tilde 1 \tilde 2}-i f_{\tilde 1 2}+if_{ 1 \tilde 2}-A)  \,.
\ee
Here $\tilde{C}^{(0)}_{12}$ denotes the LO central emission vertex of ref.~\cite{Bartels:2011ge}, with the $\nu_r$ replaced with the rapidities $u_r$,
{\small
\begin{align}
\label{eq:C0}
 \!\! \! \! \tilde C^{(0)}_{12} &\!=\! \frac{\Gamma  \bigl( 1 \! - \! i u_1 \!-\! \tfrac{n_1}{2} \!\bigr)\Gamma \bigl(1 \!+\! i u_2 \!+\! \tfrac{n_2}{2}\! \bigr) \Gamma \bigl(i u_1 \!-\! i u_2 \!-\! \tfrac{n_1-n_2}{2} \! \bigr)}{\Gamma \bigl(i u_1 \!-\! \tfrac{n_1}{2} \! \bigr) \Gamma \bigl(- i u_2 \!+\! \tfrac{n_2}{2}\! \bigr) \Gamma \bigl(1 \!-\! i u_1 \!+\! i u_2 \!-\! \tfrac{n_1-n_2}{2} \! \bigr)}.
\end{align}
}
The exponential factor and $Z_{12}$ in (\ref{eq:C_all_loop}) are obtained by analytically continuing the corresponding functions appearing in the pentagon transition \footnote{The measure is similarly obtained and is given by $\frac{d\nu_r}{2\pi} \tilde{\Phi}_r = \frac{du_r}{2\pi}\frac{g^2(x_r^+x_r^- -g^2)\exp({A-f_{rr}+f_{\tilde r \tilde r}})}{x_r^+ x_r^- \sqrt{(x_r^+ x_r^+ -g^2)(x_r^- x_r^- - g^2)}}$, see \cite{Basso:2014pla}.}. The functions $f_{rs}$ are given by
\be
\label{fhatfns}
f_{rs} = 4 \kappa(u_r,n_r) \cdot \mathbb{Q} \cdot \mathbb{M} \cdot \kappa(u_s,n_s)\,, \\
\ee
similarly $f_{\tilde r s}$ ($f_{\tilde r \tilde s}$) for $\kappa_r \to \tilde \kappa_r$ (and $\kappa_s \to \tilde \kappa_s$),
in terms of the same sources $\kappa, \tilde \kappa$ appearing in~\eqref{eq:kappa}. The constant $A$ is given by
\begin{equation}
A = 2 \int_0^\infty \frac{dt}{t} \frac{1-J_0(2gt)^2}{e^t-1} - {\pi^2}\Gamma\,.
\end{equation}
For $Z_{12}$ we have
\be\label{eq:Zfactor}
Z_{12} = \sqrt{\frac{(x_1^- x_2^- - g^2)(x_1^+ x_2^+ - g^2)}{(x_1^+ x_2^- - g^2)(x_1^- x_2^+ - g^2)}}\,,
\ee
where we introduce the Zhukowski variables
\be
\label{Zhukowski}
x_r^\pm = x(u_r \pm i\tfrac{n_r}{2})\,, \,\,\, x(u_r)=\tfrac{1}{2}(u_r+\sqrt{u_r^2-4g^2})\,.
\ee

The quantity $k_{12}$ in (\ref{eq:C_all_loop}) collects all the factors we have not addressed so far, and is a priori unknown. Nevertheless, it is constrained by the exact bootstrap condition \eqref{eq:R7hhh_bootstrapcond3} to be free of poles at $u_r=u_s$, and this condition also fixes the value of $k_{12}$ at $(u_2,n_2)=(u_1,n_1)$ to be
\begin{align}\label{eq:h12_constraint}
k_{12|(u_2,n_2)=(u_1,n_1)} &\,= \frac{x^+ x^-}{u_1^2+\tfrac{n_1^2}{4}}e^{{i}\pi \omega_1}\\
\nonumber&\, = e^{2\int_0^\infty\frac{dt}{t} \left[1-J_0(2gt)\right]\cos(u_1t)e^{-\frac{n_1}{2} t}+{i}\pi \omega_1} \,.
\end{align}

There could be many functions $k_{12}$ that satisfy~\eqref{eq:h12_constraint}, but there is a particularly simple solution where $k_{12}$ takes a factorised form,
\be
k_{12} = k_1\check{k}_2\,, \quad \check{k}(u,n) = k(-u,-n)\,.
\ee
This form is motivated by the fact that it reproduces the perturbative expansion of the same quantity to three loops, extracted from the corresponding seven-particle maximally helicity violating (MHV) amplitude \cite{Drummond:2014ffa} with the method described in \cite{DelDuca:2018hrv}. We conjecture that this minimal form persists to all orders in perturbation theory. Inserting the factorised form into~\eqref{eq:h12_constraint}, we find
\be
k_1 =  \sqrt{\frac{x^+ x^-}{u_1^2+\tfrac{n_1^2}{4}}}\,e^{\tfrac{i}{2}\pi \omega_1}\,k_{\textrm{o}1}\,,\quad \check{k}_{\textrm{o}1} = k_{\textrm{o}1}^{-1}\,.
\ee
The remaining freedom $k_{\textrm{o}1}$ can be determined by solving the exact bootstrap condition  \eqref{eq:R7hhh_bootstrapcond2} order-by-order in perturbation theory. We observe empirically that the perturbative expansion of $k_{\textrm{o}1}$ is consistent with an exponential form for $k_{\textrm{o}1}$ very reminiscent of~\eqref{eq:h12_constraint},
\be\label{eq:hodd}
k_{{\rm o}1} = e^{i\int_0^{\infty}\frac{dt}{t}\frac{(J_0(2gt)-1)(e^t+1)}{(e^t-1)}\sin(u_1t)e^{-\frac{n_1}{2}t}+\pi(u_1-\nu_1)}.
\ee

This concludes our conjecture for the all-order structure of $\cR_N$ in MRK. In fact, the dispersion integral \eqref{conj} is valid also at finite coupling, and so is the central emission block \eqref{eq:C_all_loop}, for all integer angular momenta $n_r$ different from zero. As noted in \cite{Basso:2014pla}, a subtlety that appears when $n_r=0$ is that one needs two sheets in the rapidity $u_r$ in order to cover the entire real $\nu_r$ line, with the expressions~\eqref{eq:C0}-\eqref{eq:hodd} only covering the interval $|\nu_r|\ge \tilde\nu_r=\nu(u_r=2g)$ (this is not an issue at weak coupling, where we can express all building blocks as functions of $\nu_r$ directly).
Covering also the $|\nu_r|<\tilde\nu_r$ interval would additionally serve as a starting point for analyzing the strong-coupling limit, and making contact with the string-theoretic description of the same regime \cite{Bartels:2014ppa}.

The perturbative expansion of all quantities entering (\ref{conj}) is simple to obtain \cite{Basso:2013aha,Basso:2014nra,Drummond:2015jea}, since at fixed order only a finite number of components of the vectors (\ref{eq:kappa}) contribute. The coefficients of the perturbative expansion take a very special form; the ratio to their leading-order contribution is always a polynomial in the following \emph{FM building blocks}, first introduced in~\cite{Dixon:2012yy,DelDuca:2018hrv},
\begin{align}\label{eq:FMbuildingblocks}
V_i &= \frac{i \nu_i}{\nu_i^2 + \frac{n_i^2}{4}}\,, \quad
N_i = \frac{n_i}{\nu_i^2 + \frac{n_i^2}{4}}\,,\quad D_i = -i \frac{\partial}{\partial \nu_i}\,, \notag \\
E_i &= \psi \bigl( 1 + i \nu_i + \tfrac{|n_i|}{2} \bigr) + \psi \bigl( 1 - i \nu_i + \tfrac{|n_i|}{2} \bigr)  \notag \\ &\quad - 2\psi(1) -\frac{1}{2} \frac{|n_i|}{\nu_i^2 + \frac{n_i^2}{4}} \,, \notag \\
M_{ij} &= \psi \bigl(i \nu_{ij} - \tfrac{n_{ij}}{2}\bigr) +  \psi \bigl(1 - i \nu_{ij} - \tfrac{n_{ij}}{2}\bigr)- 2 \psi(1)\,,
\end{align}
where $\nu_{ij} = \nu_i - \nu_j$, $n_{ij}= n_i - n_j$ and $\psi(z)=\partial_z \ln\Gamma(z)$ is the digamma function.

We implement the general expansion of $\tilde C^+_{12}$, and provide explicit results through five loops,
as ancilliary material with the arXiv version of the paper. As independent checks, we have verified that by inserting it to the dispersion integral~\eqref{conj} and evaluating, we find perfect agreement for the imaginary part of the four-loop seven-particle MHV symbol~\cite{Dixon:2016nkn}, as well as for the two-loop MHV amplitude at any multiplicity~\cite{Bargheer:2015djt,DelDuca:2018hrv}. More details on the integral evaluation step are provided in the next section.

\section{Analytic loop amplitudes in MRK}

In this section we provide the last ingredient needed to compute amplitudes from the dispersive representation in eq.~\eqref{conj}, and we discuss
 how the integrals can be efficiently performed in terms of the relevant class of functions in the limit, known as \emph{single-valued MPLs} (SVMPLs) \cite{BrownSVMPLs,Brown_Notes,DelDuca:2016lad}. As an application, we will give for the first time a proof of the principle of uniform and maximal transcendentality in MRK:
\begin{quote}
An $L$-loop gluon amplitude in MRK in planar $\cN=4$ SYM is a linear combination of products of $\log\tau_i$, SVMPLs, zeta values and powers of $2\pi i$ of uniform weight $2L$, for any helicity configuration and any number of legs.
\end{quote}
The proof is constructive, thereby providing an important algorithm to compute any scattering amplitude in MRK order by order in the coupling, as we now sketch. For $N=6$ gluons, similar proofs for the relevant classes of functions in the collinear and LLA multi-Regge limit have appeared in~\cite{Dixon:2012yy,Papathanasiou:2013uoa,Papathanasiou:2014yva} and \cite{Pennington:2012zj,Broedel:2015nfp} respectively, see also \cite{DelDuca:2016lad} for an extension of the latter to any $N$.

We start by noting that at order $\cO(g^2)$, the MHV amplitude will be the $(N-5)$-fold FM transform of the
\emph{vacuum ladder},
\begin{align}\label{eq:vacladder}
\varpi = \prod_{r=1}^{N-5} \frac{1}{\nu^2_r+\frac{n^2_r}{4}}  \prod_{r=1}^{N-6} {\tilde{C}_{r,r+1}^{(0)}}  \,.
\end{align}
Letting $\cF[X_r]$ denote the FM transform of $X_r$, we have in particular that $\cF[\varpi] = \delta/(4\pi)$ ,
with $\delta$ as in (\ref{conj}) being of uniform weight one.

At higher loops, the integrand will be a product of \eqref{eq:vacladder} with sums of polynomials of the FM building blocks \eqref{eq:FMbuildingblocks}. If we assign weight 1 to them, and given that the polynomial coefficients are $\mathbb{Q}$-linear combinations of Riemann zeta values $\zeta_n=\zeta(n)$, whose weight is $n$, then we observe that these polynomials have uniform transcendental weight. In other words, we see that the all-order formul\ae\ obtained from integrability imply the principle of uniform and maximal transcendentality in FM space.

To go to momentum space, we then make use of the FM transform's property to map products to convolutions,
\begin{align}
\cF[f\cdot g] = \cF[f]*\cF[g]\,,
\end{align}
where
\begin{align}
(F*G)(z) = \int \frac{d^2w}{|w|^2} F(w)G\left(\frac{z}{w}\right) \,.
\end{align}
Every higher-loop amplitude in MRK can thus be built iteratively by convolving the vacuum ladder~\eqref{eq:vacladder} with a finite number of FM building blocks~\eqref{eq:FMbuildingblocks}. While the evaluation of the convolution integral seems a daunting task, it was shown in~\cite{Schnetz:2013hqa} (see also~\cite{DelDuca:2016lad}) that, in the case where the integrand only involves rational functions and SVMPLs, the integral can easily be evaluated in terms of residues.

The proof now proceeds by induction: Assume we have a pure linear combination of SVMPLs of uniform weight. We will show that convolution with any FM building block raises the weight by 1 and preserves purity. This justifies our assignment of weight 1 to the building blocks, and implies that all MHV amplitudes in MRK satisfy the principle of uniform  and maximal transcendentality.

More concretely, assume that $f(z)$ is a pure linear combination of SVMPLs of uniform weight $n$ and let
\begin{align}\label{cK}
\cK(z) = |z|^2 \sum_{i,j} \frac{a_{ij}}{(z-\alpha_i)(\zb - \beta_j)} \,,
\end{align}
with $a_{ij}, \alpha_i, \beta_j \in \mathbb{Q}$. One can show using Stokes' theorem~\cite{Schnetz:2013hqa} that $(f\ast\cK)(z)$ is again pure and has uniform weight $n+1$. The FM transform of the building blocks $E_r,N_r,V_r$ match the form in~\eqref{cK}~\cite{DelDuca:2016lad,DelDuca:2018hrv,Marzucca:2018ydt}
\begin{align}
\cF\left[E_r\right] &= - \frac{z_r + \zb_r}{2 |1-z_r|^2} \,, \\
\cF\left[V_r\right] &= \frac{2 - z_r - \zb_r}{2|1-z_r|^2} \,, \\
\cF\left[N_r\right] &=  \frac{z_r - \zb_r}{|1-z_r|^2} \,.
\end{align}
Hence they raise the weight of the function they are convolved with by $1$. We may similarly show that the same holds true for the derivative $D_r$, by using integration by parts to let it act on the factor $|z_r|^{2 \nu_r}$ in the definition of the FM transform,
\begin{align}
\cF[D_r X_r] = -\log|z_r|^2\, \cF[X_r] \,.
\end{align}
Finally, let us note that the FM building block $M_{rs}$ obeys
\begin{align}
M_{rs} &= D_{r} \log(\tilde{C}_{rs}^{(0)})+E_r+V_r\\
&= - D_{s} \log(\tilde{C}_{rs}^{(0)})+E_s-V_s\,.
\end{align}

This allows us to shift occurrences of $M_{rs}$ in its FM transform with the vacuum ladder to either end,
\begin{align}
 \cF[\varpi M_{r\,{r+1}}] &= \cF[\varpi M_{r-1\,r}]+\cF[D_r\varpi] \\
 \cF[\varpi M_{12}] &= \cF[\varpi E_1]-\cF[\varpi V_1] + \cF[D_1 \varpi] \,,
\end{align}
and in this manner replace it by a combination of $E,V,D$. Hence, $M_{rs}$ raises the weight of the integral by $1$ as well.  Finally, our proof may be immediately  extended to non-MHV amplitudes as well. The latter can be obtained by convoluting MHV amplitudes with the helicity-flip kernel $I^-$, and the only difference is that at LO the latter does not raise the weight, and it does not preserve the purity of the function~\cite{DelDuca:2016lad}. We therefore conclude that non-MHV amplitudes have the same weight as their MHV counterparts, but are no longer pure functions.

\section{Conclusions}

We have presented a dispersion integral for all gluon amplitudes of arbitrary multiplicity, helicity configuration in MRK. By combining our results with~\cite{DelDuca:2016lad} we obtain an efficient algorithm to evaluate any scattering amplitude in MRK, for any number of loops or legs, and for arbitrary helicity configurations.

We believe that our results, while complete for the sector of planar $\mathcal{N}=4$ super Yang-Mills theory that we have studied, should serve as the basis for many future generalisations in various directions. Firstly it should be straightforward to include the fermions and scalars into our expression, or to consider more general Mandelstam regions~\cite{Bargheer:2015djt,Bargheer:2016eyp,DelDuca:2018raq,Bargheer:2019lic}. We believe that a similar structure will survive for general gauge theories, at least in the planar limit, though the details will differ because in general dual conformal symmetry is broken. It would be very interesting to understand how the form of the amplitude generalises beyond the planar limit.

\section*{Acknowledgements}

We would like to thank Jochen Bartels and Benjamin Basso for comments on the manuscript. This work was supported in part by the ERC Consolidator grant 648630 IQFT and the
ERC Starting grants 637019 MathAm and 804286 UNISCAMP, as well as the U.S.Department of Energy (DOE)
under contract DE-AC02-76SF00515.

\appendix

\section{Appendix: BES kernel and BFKL sources}
\label{App-BCS}

The sources $\kappa$ and $\tilde{\kappa}$ are infinite-dimensional vectors with $j$-th component given by
\begin{align}\label{eq:kappa}
\kappa_j(u_r,n_r) &= - \int_0^\infty \frac{dt}{t} \frac{J_j(2gt)}{e^t-1} \phi_j(t;u_r,n_r) \,,
\end{align}
and similarly for $\tilde{\kappa}$ with $\phi$ replaced by $\tilde{\phi}$. Here $J_j(x)$ denote Bessel functions of the first kind, and we have
\begin{align}
\phi_j &\!=\! \tfrac{1}{2}\!\bigl[e^{\frac{t(1+(-)^j)}{2}} \!-\! (-)^j e^{\frac{t(1-(-)^j)}{2}}\bigr]\! \cos (u_rt) e^{-\frac{n_rt}{2}} \!-\!J_0(2gt), \notag \\
\tilde{\phi}_j &\!=\!  \tfrac{1}{2}\!\bigl[e^{\frac{t(1+(-)^j)}{2}} \!+\! (-)^j e^{\frac{t(1-(-)^j)}{2}}\bigr]\! \sin (u_rt) e^{-\frac{n_rt}{2}}\,.
\end{align}
The matrices $\mathbb{Q}$ and $\mathbb{M}$ in (\ref{eq:nutou_weak}) are given by
\begin{align}
&\mathbb{Q}_{ij} = \delta_{ij}(-1)^{i+1}i\,, \qquad \quad \mathbb{M} = (1+ \mathbb{K})^{-1}\,, \notag \\
& \mathbb{K}_{ij} = 2j (-1)^{j(i+1)} \int_0^{\infty} \frac{dt}{t} \frac{J_i(2gt)J_j(2gt)}{e^t-1}\,,
\end{align}
with the latter being simply the kernel of the Beisert-Eden-Staudacher equation \cite{Beisert:2006ez} as reformulated in \cite{Benna:2006nd}.

\bibliographystyle{apsrev4-1}
\bibliography{refs}

\end{document}